\documentstyle[preprint,aps]{revtex}
\def\bra#1{\langle #1 \,\vert}
\def\ket#1{\vert\, #1 \rangle}

\def\lag{\langle}
\def\rag{\rangle}
\def\Tr{{\rm Tr}}

\def\d{\mbox{d}}
\def\e{\mbox{e}}
\def\sign{\mbox{sign}}

\begin{document}
\title{Axial form factor in the Chiral Quark Soliton Model}
\author{T. Watabe\thanks{watabe@baryon.tp2.ruhr-uni-bochum.de},
Chr.V. Christov\thanks{Permanent address:
Institute for Nuclear Research and Nuclear Energy, Sofia,
Bulgaria}\thanks{christov@neutron.tp2.ruhr-uni-bochum.de}
and K. Goeke\thanks{goeke@hadron.tp2.ruhr-uni-bochum.de}}
\address{ Institut f\"{u}r Theoretische Physik II, Ruhr-Universit\"at
Bochum, D-44780 Bochum, Federal Republic Germany}
\maketitle
%
%%%%%%%%%%%%Abstract%%%%%%%%%%%%%%%%%%%%%%%%%%%%%%%%%%%%%%%%%%%%%%%%
%
\vspace {-6cm}
\baselineskip=0.4cm
\begin{flushright}{RUB-TPII-3/95 \\ \quad \\ April,95}
\end{flushright}
\vspace {6cm}
\tighten
\begin{abstract}
We calculate the axial form factor in the
chiral quark soliton (semibosonized Nambu - Jona-Lasinio) model  using the
semiclassical quantization scheme in
the next to leading order in angular velocity. The obtained axial form
factor is in a good absolute (without additional scaling) agreement
with the experimental data. Both the value at
the origin and the $q$-dependence of the form factor as well as the
axial m.s.radius are fairly well reproduced.
\end{abstract}
\draft
\newpage
%%%%%%%%%%%%Main%%%%%%%%%%%%%%%%%%%%%%%%%%%%%%%%%%%%%%%%%%%%%%%%%%%%
%
Recently, including the $1/N_c$ rotational corrections (next to leading order)
in the semiclassical quantization scheme, a natural solution for the problem of
 strong underestimation of the axial-vector coupling constant $g_A$ in the
leading order in the semibosonized Nambu - Jona-Lasinio (chiral quark soliton)
model
has been found~\cite{Wakamatsu93,Christov94,Blotz93,Christov95}. These
corrections lead to an enhancement~\cite{Christov94} of order $(N_c+2)/N_c$
for $g_A$ and improve considerably the agreement with the experiment.
Apparently one should expect that the $1/N_c$ corrections will not be small
also for the axial form factor and it can change significantly the existing
leading order results~\cite{Meissner91}. It should be noted that the latter
suffer from the fact that
$g_A=G_A(0)$ is strongly underestimated which does not allow
for an absolute agreement with experiment.  It is usually
assumed~\cite{Meissner91} that the axial form factor is scaled in the same
way like the axial vector
coupling constant $g_A$. This is, however,  not necessarily true since the
axial
form factor contains a second contribution, which does not contribute to
$g_A=G_A(0)$. Also in the calculation~\cite{Meissner91} a low
constituent mass $M=363$ MeV is used, which differs from the value
$M=420$ MeV extracted from a overall fit~\cite{Christov95ff} to the static
properties as well as to the electromagnetic form factors of the nucleon in
the model.
Therefore, it is worth to evaluate the axial form factor in the chiral
quark soliton model using the semiclassical quantization scheme with the
next to leading $1/N_c$ rotational corrections included, which is
the aim of the present work.

We start with the general decomposition of the matrix element of the axial
current $A^\mu_a=\Psi^\dagger\gamma^0\gamma^\mu \gamma_5\frac {\tau_a}2
\Psi$ in terms of the corresponding form factors:
\begin{equation}
\langle  N(p^\prime,\xi^\prime)|
\Psi^\dagger\gamma^0 \gamma^\mu \gamma_5\frac {\tau_a}2 \Psi
|N(p,\xi)\rangle
= {\bar u}(p^\prime,\xi^\prime)\Bigl[ G_A(q^2)\gamma^\mu + {G_p(q^2)\over
2M_N}q^\mu \Bigr]\gamma_5 \frac{\tau_a}2 u(p,\xi)\,, \label{GA-definition}
\end{equation}
where $\quad q=p'-p\quad$ and $\xi$ stands for spin and isospin. After some
standard manipulations we can express the axial form factor as
\begin{equation}
G_A(q^2) = 3 \ \frac{M_N}{E} \ \int d^3x \ e^{i \vec{q} \vec{x}} \ [
\lag p \uparrow | A^3_3(x) | p \uparrow \rag
- {q_3  q_i\over q^2} \lag p \uparrow | A^i_3(x) | p \uparrow \rag
] ,
\label{GA-ff}
\end{equation}
where $M_N$ and $E$ are the nucleon mass and energy
$E=\sqrt{M_N^2+\vec{q}^2/4}$, respectively, and $| p \uparrow \rangle$ is a
proton state of spin up. It should be noted that eq.(\ref{GA-ff}) contains
matrix elements of the space components of the axial current $A^i_a$, and
apparently only the first term in eq.(\ref{GA-ff}) contributes to $g_A$.

For the evaluation of the matrix element of the axial current we use the
simplest SU(2)-version of the chiral quark soliton model with up and down
quarks, degenerated in mass. It is based on a semibosonized Nambu
Jona-Lasinio lagrangean~\cite{Nambu61,Eguchi76}:
\begin{equation}
{\cal L} = \bar{\Psi}(-i \gamma^{\mu} \partial_{\mu}
+ m_0 + M U^{\gamma_5}) \Psi\,,
\label{lagrangean}
\end{equation}
which includes auxiliary meson fields
\begin{equation}
U(\vec x) = \e^{i \vec{\tau} \cdot \vec{\pi}(\vec x) / f_{\pi}}\,,
\label{u}
\end{equation}
constrained on the chiral circle. The model is non-renormalizable and
one needs a cutoff to make it finite. This cutoff $\Lambda$ and the current
quark mass $m_0$ are treated as parameters of the model and are both fixed in
the
mesonic sector to reproduce the physical pion mass $m_\pi$ and the pion
decay constant $f_\pi$. The third model parameter, the constituent quark
mass $M$, can be related to the empirical value of the quark condensate
but it still leaves a broad range for $M$. Actually, in order
to obtain an overall good description of the baryonic
properties a value around $420$ MeV has to be used~\cite{Christov95ff}.

In the model the baryons appear as a bound state of $N_c$ (number of colors)
valence quarks coupled to the polarized Dirac sea. Since the model lacks
confinement the proper way to describe the nucleon is to
consider~\cite{Diakonov88} a correlation function of two $N_c$-quark
currents with nucleon quantum numbers at
large euclidean time-separation. Accordingly, the nucleon matrix element of the
axial current $\Psi^\dagger\gamma^0\gamma^\mu \gamma_5{\tau_a\over 2}\Psi$,
is represented by an euclidean functional integral~\cite{Diakonov88} with
lagrangean (\ref{lagrangean}):
\begin{eqnarray}
&&\langle N', {\vec p}' | \Psi^\dagger(0)\gamma^0
\gamma^\mu \gamma_5{\tau_a\over 2}\Psi(0) | N, {\vec p} \rangle
=\lim_{ T\to \infty}
\frac 1Z \int \d^3 x \d^3y
\e^{ - i{\vec p}' {\vec x}' + i{\vec p} {\vec x}}\nonumber \\
&&\times\int{\cal D}U \int {\cal D}\Psi \int {\cal D}\Psi^\dagger
J_{N^\prime}(\vec x^\prime,T/2) \> \Psi^\dagger(0) \gamma^0\gamma^\mu
\gamma_5{\tau_a\over 2} \Psi(0) J_N^\dagger(\vec x,-T/2) \e^{- \int d^4 z
\Psi^\dagger D(U) \Psi }\,.
\label{form-factor-integral}
\end{eqnarray}
Current $J_N$ is a composite $N_c$ quark operator~\cite{Ioffe81} with nucleon
quantum numbers $J J_3, T T_3$:
\begin{equation}
J_N(\vec x,t) = \frac{1}{N_c !} \
\varepsilon^{\beta_1\cdots \beta_{N_c}}
\Gamma^{ f_1\cdots f_{N_c}JJ_3,TT_3} \
\Psi_{\beta_1 f_1}(\vec x,t) \cdots \Psi_{\beta_{N_c} f_{N_c}}(\vec
x,t)\,.
\label{jn}
\end{equation}

For the evaluation of the path integral
in eq.(\ref{form-factor-integral})
we follow the line of ref.~\cite{Christov94,Christov95ff}. Here we will
only sketch the derivation.
Integrating out the quarks in (\ref{form-factor-integral}) it is easy to
see that the result is naturally split in valence and sea parts.  After that
we integrate over the meson fields $U$ in saddle point approximation  -- large
$N_c$ limit. It leads to a stationary localized meson configuration (soliton)
of hedgehog structure
\begin{equation}
\bar U(x)\,=\,\e^{i\vec{\tau}\cdot\hat x\,P(x)}\,,
\label{HEDGEHOG}\end{equation}
which minimizes the effective action
\begin{equation}
\Tr\log D(U)=\Tr\log[\partial_\tau+h(U)]\,,
\label{DIRAC}\end{equation}
where the one-particle hamiltonian $h$ is given by
\begin{equation}
h(U)=\frac {\vec{\alpha}\cdot\vec{\nabla}}i+\beta M U^{\gamma_5}+\beta
m_0\,.
\label{HAMIL}\end{equation}
Since the hedgehog soliton field configuration $\bar U(x)$ does not
preserve the spin and isospin, as a next step we make use of the rotational
zero modes to quantize it~\cite{Diakonov88}. It is done assuming a rotating
meson hedgehog fields of the form
\begin{equation}
U({\vec x},\tau)=R(\tau)\,\bar U({\vec x})\, R^\dagger(\tau)\,
\label{ROTHEDG}\end{equation}
with $R(\tau)$ being a time-dependent rotation SU(2) matrix in the isospin
space. It means that even in leading order in $N_c$ one has to go beyond the
saddle point approximation extending the path integral in
eq.(\ref{form-factor-integral}) over all fields of the form (\ref{ROTHEDG}) --
a path integral over $R$.
It is easy to see that for such an ansatz one can transform the effective
action \begin{equation}
\Tr\log D(U)=\Tr\log(D(\bar U)+i\Omega)
\label{ROTEFA}\end{equation}
in order to separate the angular velocity matrix:
\begin{equation}
\Omega=-iR^\dagger(\tau)\dot R(\tau)=\frac 12\Omega_a\tau_a\,.
\label{OMEGA}\end{equation}
The dot stands for the derivative with respect to the euclidean time $\tau$.
Similarly, the quark propagator in the background meson field $U$ can be
rewritten as
\begin{equation}
\bra{x}\,\frac 1{D(U)}\, \ket{x^\prime} \,=\,\bra{x}\,R(\tau)\,\frac 1{D(\bar
U)+i\Omega}\,R^\dagger(\tau^\prime)\, \ket{x^\prime} \,.
\label{Eq10b}\end{equation}

Since the angular velocity is quantized according to the canonical
quantization rule, it appears as $\Omega_a\sim \frac 1{N_c}$. This allows
one to consider $\Omega$ as perturbation and to evaluate any observable as a
perturbation series in $\Omega$ which is actually an expansion in $\frac
1{N_c}$. Actually, we make essentially use of the expansions
\begin{equation}
N_c \Tr \log [D(\bar U)+i\Omega]
= N_c \Tr \log [D(\bar U)]
+\Theta^{sea} \int d\tau \,\Omega^2_a + \ldots\,,
\label{moment-of-inertia-definition}
\end{equation}
and
\begin{equation}
\frac 1{D(\bar U)+i\Omega}=\frac 1{D(\bar U)}
-\frac 1{D(\bar U)} i\Omega \frac1{D(\bar U)}
+\ldots\,.
\label{Dexpansion}
\end{equation}
up to terms quadratic in $\Omega$ to arrive at
a functional integral over the time dependent orientation matrices $R(\tau)$
with an action quadratic in angular velocities~\cite{Diakonov88}.
In large $N_c$ limit, the latter corresponds to the hamiltonian of the
quantum spherical rotator:
\begin{equation}
H_{rot} = J_a^2/(2\Theta)\,,
\label{h-rot}
\end{equation}
where $J_a$ is the spin operator of the nucleon and
\begin{equation}
\Theta=\Theta^{sea}+\Theta^{val}\sim N_c
\label{Theta}
\end{equation}
is the total moment of inertia~\cite{Diakonov88,Reinhardt89}, including
both the valence and the sea quark
contributions. It means that despite of the fact that in general the
path integral over $R$
runs over all possible trajectories, in large $N_c$ limit the main
contribution comes from trajectories close to those of the quantum rotator
with hamiltonian~(\ref{h-rot}). According to eq.~(\ref{h-rot}) the quantization
rule (in euclidean space-time) is given by
\begin{equation} \Omega_a \to -i\frac{J_a}{\Theta} \,.
\label{quantization-rule}
\end{equation}
Due to the collective path integral over $R$ the order of the not-commuting
collective operators
\begin{equation}
\Omega_a(R(\tau))=-i\Tr \bigl(R^\dagger(\tau)\dot R(\tau)\tau_a\bigr)\quad
\mbox{and}\quad
D_{ab}(R(\tau))= {1\over 2}\Tr \bigl(R^\dagger(\tau)\tau_a
R(\tau)\tau_b\bigr)\,,
\label{colloper}
\end{equation}
is strictly fixed by the time ordering\footnote{
The details of this procedure can be found in ref.\cite{Christov95ff}.}.
For given spin $J,J_3$ and isospin $T,T_3$ the
spin-flavor structure of the nucleonic solution can be expressed through the
Wigner $D$ function \begin{equation}
|N,T_3J_3 \rangle (R) = (-1)^{T+T_3}\sqrt{2T+1}\,D^{T=J}_{-T_3,J_3}(R)\,.
\label{D-functions}
\end{equation}

In the above scheme, the matrix element of the space components of the axial
current $A^k_3$ includes leading order terms~$\sim \Omega^0$ as well as
next to leading order ones $\sim\Omega$ ($1/N_c$). In Minkowski
space-time it has the following structure:
\begin{eqnarray}
&&\bra{N} A^k_3(\vec x)\ket{N}=-N_c
\Biggl\{\Bigl(\Phi^\dagger_{val}({\vec x})
\,\gamma^0\gamma^k\gamma_5\tau_3\,\Phi_{val}({\vec
x})\Bigr)-\sum\limits_n {\cal R}^{\Omega^0}_\Lambda(\epsilon_n)
\Bigl(\Phi^\dagger_n({\vec x}) \gamma^0 \gamma^k\gamma_5 \tau_b
\Phi_n({\vec x})\Bigr)
\nonumber\\
&&+\frac i{2\Theta}\varepsilon^{cb3} \Biggl[\sum\limits_{n\neq
val}\sign({\epsilon_n})
\frac {\Bigl(\Phi^\dagger_{val}({\vec
x})\gamma^0\gamma^k\gamma_5\tau_b\Phi_n({\vec
x})\Bigr)\bra{n}\tau_c\ket{val}}{\epsilon_n -
\epsilon_{val}}\nonumber\\
&&-\sum\limits_{n,m}{\cal
R}^{\Omega^1}_\Lambda(\epsilon_n,\epsilon_m)\Bigl(\Phi_m^{\dagger}({\vec x})
\gamma^0 \gamma^k \gamma_5\tau_b\Phi_n({\vec x})\Bigr)
\bra{n}\tau_c\ket{m}\Biggl]\Biggr\}\,.
\label{MEAC}\end{eqnarray}
Here $\Phi_n$ and $\epsilon_n$ are the eigenfunctions and the eigenvalues of
the hamiltonian (\ref{HAMIL}). In eq.(\ref{MEAC}) first and
third terms are
valence quark contributions in leading and next to leading order in
angular velocity, respectively. They are finite and do not need any
regularization. The other two
terms represent the divergent Dirac sea part and need regularization.
The regularization functions
${\cal R}_\Lambda^{\Omega^0}$, ${\cal R}^{\Omega^1}_\Lambda$ can be
found in refs.\cite{Christov94,Christov95ff}.

Inserting the result (\ref{MEAC}) in eq.(\ref{GA-ff}), after some
straightforward calculations we obtain for the axial factor
\begin{equation}
G_A(q^2) = \frac{M_N}{E} \ \int r^2 dr \
[j_0(qr) A_0(r) -j_2(qr) A_2(r)] \,,
\label{mtfmGA}
\end{equation}
where both densities $A_0(r)$ and $A_2(r)$, split in valence and sea parts,
contain leading and next to leading order terms in angular velocity
$\Omega$:
\begin{eqnarray}
A_{0(2)}^{\Omega^0}(r) =
N_c \frac{1}{3\sqrt{3}}&& \Biggl\{
\Bigl(\Phi^\dagger_{val}(r) \|  T_{0(2)}^{\Omega^0} \|
\Phi_{val}(n)\Bigr) \nonumber\\
&&+  \frac 12
\sum_{n = all}\sqrt{2 K_n + 1}\, {\cal R}^{\Omega^0}_\Lambda(\epsilon_n)
\Bigl(\Phi^\dagger_n(r) \| T_{0(2)}^{\Omega^0} \| \Phi_n(n)\Bigr)\Biggr\} ,
\label{GA0}
\end{eqnarray}
and
\begin{eqnarray}
A_{0(2)}^{\Omega^1}(r) =
\frac{N_c}{I} \frac 1{9\sqrt{2}}&&\Biggl\{
\sum_{n \neq val} \frac{1}{\epsilon_n - \epsilon_{val}} \
\Bigl(\Phi^\dagger_{val}(r) \| T_{0(2)}^{\Omega^1}  \| \Phi_n(r)\Bigr)
\langle val \| \tau^{(1)} \| n \rangle\nonumber\\
&&+
\frac{1}{2} \sum_{n,m = all} {\cal
R}^{\Omega^1}_\Lambda(\epsilon_n,\epsilon_m)
\Bigl(\Phi^\dagger_n(r) \| T_{0(2)}^{\Omega^1} \|\Phi_m(r)\Bigr)
\lag n \| \tau^{(1)} \| m \rag\Biggr\}\,.
\label{GA1}
\end{eqnarray}
The tensors  $T_{0(2)}^{\Omega^{0(1)}}$ are defined as
\begin{equation}
T_{0}^{\Omega^{0(1)}} =
[\sigma^{(1)} \otimes \tau^{(1)}]^{(0(1))} \,,
\label{tens00}
\end{equation}
and
\begin{equation}
T_{2}^{\Omega^{0(1)}} =
\sqrt{2\pi} \ [ \ [Y_2 \otimes \sigma^{(1)}]^{(1)} \otimes
\tau^{(1)}]^{(0(1))} \,.
\label{tens01}
\end{equation}

The solitonic solution $\bar U$ is found by solving
numerically the corresponding equations of motion in an iterative
self-consistent procedure~\cite{Meissner94}. To this end we use the method
of Ripka and Kahana~\cite{Ripka84} for solving the eigenvalue problem in a
finite quasi--discrete basis.

We also calculate the axial m.s. radius $\lag r^2 \rag_A$ given by
\begin{equation}
\lag r^2 \rag_A =
- \ \frac{6}{G_A(0)} \ \frac{\mbox{d} G_A(q^2)}{\mbox{d} q^2}
\Bigg\vert_{q^2=0}
= \frac{1}{G_A(0)} \int r^4 \mbox{d}r \ [A_0(r)+\frac{2}{5}A_2(r)]
+\frac{3}{4M_N^2} \,,
\label{sqradi}
\end{equation}
and for convenience, parameterize also the calculated form factors using
a dipole fit:
\begin{equation}
\frac{G_A(q^2)}{G_A(0)} =
\left( 1 + \frac{q^2}{M_A^2} \right)^{-2} \,.
\label{dipole}
\end{equation}

Our results for the axial form factor are displayed on fig.\ref{Figr1} for
four different values of the constituent mass $M$, namely 360, 400, 420 and
440 MeV. Similar to $g_A$~\cite{Christov94} the
main contribution to the axial form factor in the leading as well as in
the next to leading order in angular velocity comes from
the valence quarks and the Dirac sea contribution is almost negligible.
Table~\ref{Tabl1} contains the corresponding values for the
dipole mass and m.s.radius as well as for the axial vector coupling
constant. Similar to the other nucleon properties~\cite{Christov95ff} a
value for constituent mass around 420 MeV is preferred. For this mass
value the theoretical curve agrees fairly well with the experimental
data without any scaling. The lowest value of 360 MeV is almost excluded.
On fig.\ref{Figr1}
we also show the leading order results. As can be seen the two groups of
curves differ in both  magnitude and slope. Indeed, the extracted
dipole masses for the axial form
factors, calculated in leading order, are larger. In particular,
for $M=360$ MeV the dipole mass in the leading order still agrees with the
experiment, whereas in the next to leading order it is out of the
experimental window. It means that a simple scaling
behavior of the axial form factor as it is assumed in
ref.~\cite{Meissner91}
is not valid. On fig.\ref{Figr2} we present the separated $A_0$- and
$A_2$-contributions to the axial form factor. As can be seen the $A_0$
part completely dominates the axial form factor at small momentum transfer.

In Table~\ref{Tabl1} we also present our results for the axial m.s.radius
calculated from eq.(\ref{sqradi}) as well as from the dipole fit (in
brackets) in comparison with the estimate from the experimental dipole
fit. As can be seen the values directly calculated from the corresponding
densities $A_0$ and $A_2$ are very close to those of the dipole fit
which is an indication that even at small $q^2$ the dipole fit is a good
approximation to the theoretical curves.

To summarize, we evaluate the axial form factor in the chiral quark soliton
model taking into account the next to leading order ($1/N_c$) rotational
corrections in the semiclassical quantization scheme. These corrections
provide not only the enhancement, needed to reproduce the experimental
value of the axial vector coupling constant, but also change the slope of the
axial form factor allowing for an almost perfect absolute (without additional
scaling) description of the experimental data. The axial
m.s.radii are also well reproduced.

\section*{Acknowledgement}

The project has been partially
supported by the Volkswagen Stiftung, DFG and COSY (J\"ulich).

%%%%%%%%%%%%Figure Captions%%%%%%%%%%%%%%%%%%%%%%%%%%%%%%%%%%%%%%%%%
%
\begin{figure}
\caption{Axial form factor in leading and next to leading order in
angular velocity in comparison with the experimental
data~\protect{\cite{Bake,Kita}}.}
\label{Figr1}
\end{figure}
\begin{figure}
\caption{Separated $A_0$- and $A_2$-contributions to the axial form
factor.}
\label{Figr2}
\end{figure}
%%%%%%%%%%%%Tables%%%%%%%%%%%%%%%%%%%%%%%%%%%%%%%%%%%%%%%%%%%%%%%%%%
%
\begin{table}
\caption{Axial properties of the nucleon, calculated in
the NJL model for four different values of the constituent
mass $M=360, 400, 420$ and $440$ MeV, compared with experimental
values. The dipole masses for the form factors calculated in both
leading and next to leading order are given.The m.s.radius obtained
from the dipole fit (in brackets) is also presented.}
\label{Tabl1}
\begin{center}
\begin{tabular}{lccccc}
 &
\multicolumn{4}{c}{{\bf Constituent quark mass $M$ [MeV]}} & \\
 & 360 & 400 & 420 & 440 & exp \\ \hline
 $M_A^{\Omega^0}$ [GeV] &
 $0.910 \pm 0.001$ & $1.004 \pm 0.001$ & $1.040 \pm 0.001$ &
 $1.072 \pm 0.001$ & \\
 $M_A^{\Omega^0+\Omega^1}$ [GeV] &
 $0.849 \pm 0.001$ & $0.956 \pm 0.001$ & $0.995 \pm 0.001$ &
 $1.031 \pm 0.001$ & $1.05^{+0.12}_{-0.16}$ \\
 $< r^2 >_A$ [${\rm fm}^2$] &
 0.70 (0.65) & 0.51 (0.51) & 0.46 (0.47) & 0.43 (0.44) &
 $0.42^{+0.18}_{-0.08}$ \\
 $g_A$ &
 1.31 & 1.24 & 1.21 & 1.18 & 1.26
\end{tabular}
\end{center}
\end{table}


\begin{thebibliography}{99}

\bibitem{Wakamatsu93}M. Wakamatsu and T. Watabe, {\it Phys.Lett.} {\bf
B312}(1993)184

\bibitem{Christov94} Chr. V. Christov, A.Blotz, K. Goeke, V. Yu. Petrov,
P. V. Pobylitsa, M. Wakamatsu, T. Watabe, {\it Phys.Lett.} {\bf
B325}(1994)467

\bibitem{Blotz93}A. Blotz, M. Praszalowicz and K. Goeke, {\it Phys.Lett.}
{\bf B317}(1993)195

\bibitem{Christov95}Chr.V.Christov, K.Goeke and P.V.Pobylitsa, {\it
Phys.Rev.} {\bf C51}(1995)

\bibitem{Meissner91} T.Meissner and K.Goeke,
{\it Z.Phys.} {\bf A339} (1991) 513

\bibitem{Meissner94} T.Meissner, A.Blotz, E.Ruiz Arriola and K.Goeke,
Baryons in Effective Chiral Quark Model with Polarized Dirac Sea, Bochum
preprint RUB-TP2-42/93, (submitted to {\it Rep.Prog.Theor.Physics})

\bibitem{Nambu61}Y.Nambu and G.Jona-Lasinio, Phys.Rev. {\bf 122} (1961) 354

\bibitem{Eguchi76} T.Eguchi, H.Sugawara,
{\it Phys.Rev.} {\bf D10} (1974) 4257;
T.Eguchi, {\it Phys.Rev.} {\bf D14} (1976) 2755

\bibitem{Diakonov88}D.I.Diakonov, V.Yu.Petrov and P.V.Pobylitsa, 21$^{st}$
LNPI Winter School on Elem.Part. Physics, Leningrad 1986; {\it Nucl.Phys.}
{\bf B306}(1988)809

\bibitem{Ioffe81}B.L.Ioffe, Nucl.Phys.{\bf B188}(1981)317;{\bf
B191}(1981)591(E)

\bibitem{Christov95ff}Chr.V. Christov, A.Z. G\'{o}rski, K. Goeke and
P. Pobylitsa, {\it Nucl.Phys.}{\bf A}(1995) (in print)

\bibitem{Reinhardt89}H.Reinhardt, {\it Nucl.Phys.} {\bf
503B}(1989)825

\bibitem{Ripka84}S.Kahana and G.Ripka, {\it Nucl.Phys.}{\bf A429} (1984) 462

\bibitem{Bake}N.J. Baker, et al, {\it Phys.Rev.} {\bf D23} (1981) 2499

\bibitem{Kita}T. Kitagaki, et al, {\it Phys.Rev.} {\bf D28} (1983) 436

\end{thebibliography}
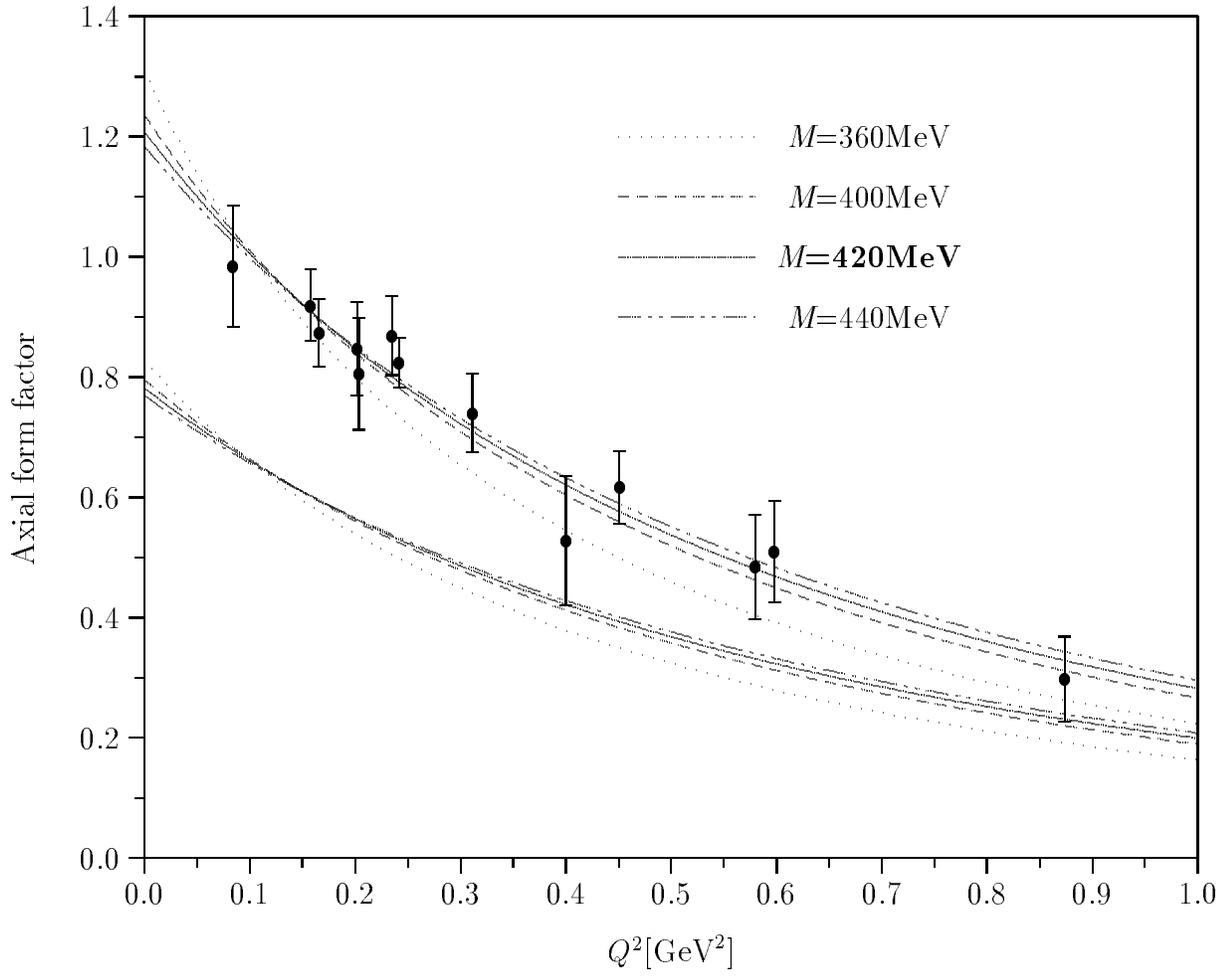
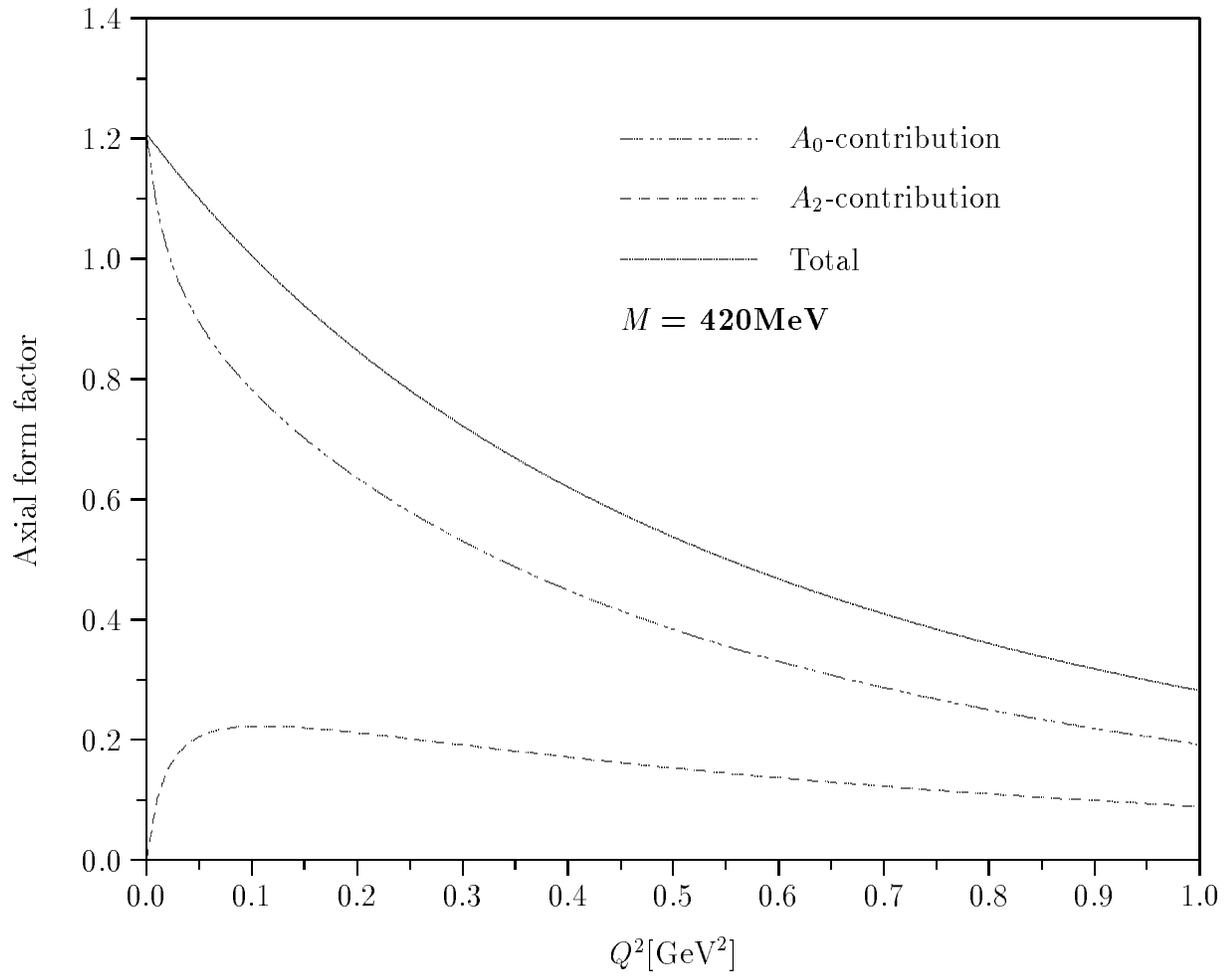
\end{document}